\documentclass{iopart}
\usepackage{amssymb}
\usepackage{epsf}
\usepackage{subeqnar}

\begin{document}

\title[Reaction-diffusion with a time-dependent reaction 
rate]{Reaction-diffusion with a time-dependent reaction rate:  
the single-species diffusion-annihilation process}

\author[L Turban]{L Turban}

\address{Laboratoire de Physique des Mat\'eriaux, Universit\'e Henri 
Poincar\'e (Nancy 1), BP~239, 
F-54506 Vand\oe uvre l\`es Nancy Cedex, France}

\ead{turban@lpm.u-nancy.fr}

%\received{}

\begin{abstract}
We study the single-species diffusion-annihilation process with a 
time-dependent 
reaction rate, $\lambda(t)=\lambda_0\,t^{-\omega}$. Scaling arguments show that 
there is a critical value of the decay exponent $\omega_{\rm c}(d)$ separating 
a reaction-limited regime for $\omega>\omega_{\rm c}$ from a diffusion-limited 
regime for $\omega<\omega_{\rm c}$. The particle density displays a mean-field, 
$\omega$-dependent, decay when the process is reaction limited whereas it 
behaves as for a constant reaction rate when the process is diffusion limited. 
These results are confirmed by Monte Carlo simulations. They allow us to discuss 
the scaling behaviour of coupled diffusion-annihilation processes in terms of 
effective time-dependent reaction rates.
\end{abstract}

\section{Introduction}
Time-dependent effective reaction rates can be introduced into mean-field 
kinetic equations in order to simulate the effect of concentration 
fluctuations below the upper critical dimension $d_{\rm c}$ where deviations 
from the standard mean-field behaviour are observed~\cite{kang84a}. In this 
work we take the opposite point of view and look how far the fluctuations in a 
reaction-diffusion process can be affected by a time-dependent reaction rate.

We study the single-species diffusion-annihilation process 
\begin{eqnarray}
&&A\,\varnothing\ \stackrel{1}{\longleftrightarrow}\ \varnothing\,A
\quad {\rm(diffusion)}\nonumber\\
&&A\,A\ \stackrel{\lambda(t)}{\longrightarrow}\ \varnothing\,\varnothing
\quad{\rm(annihilation)}\,,
\label{e1.aa}
\end{eqnarray}
where nearest-neighbour particles annihilate with a reaction rate
\begin{equation}
\lambda(t)=\frac{\lambda_0}{t^\omega}\,,
\label{e1.rr}
\end{equation}
decaying as a power of the time. The diffusion rate associated with the 
exchange of a particle ($A$) with a vacancy ($\varnothing$) is equal to 1, 
which fixes the time scale. 

When the reaction rate is constant ($\omega=0$), the kinetic equation 
$\dot{\rho}=-\lambda_0\rho^2$ leads to the mean-field asymptotic behaviour 
$\rho(t)\simeq (\lambda_0t)^{-1}$. Simple scaling 
arguments~\cite{toussaint83,kang85} indicate that, due to concentration 
fluctuations, the decay is slower in low dimensions. At time $t$, for 
$d\leq 2$ 
where the exploration is dense, a surviving particle has swept out a region 
with a linear size given by the diffusion length $\sqrt{Dt}$. Therefore, the 
volume per particle grows as $(Dt)^{d/2}$ and the particle density decays 
algebraically as
\begin{equation}
\rho(t)=C\,t^{-\alpha}\,,\qquad \alpha=\frac{d}{2}\,,\qquad d<d_{\rm c}=2\,.
\label{e1.rhodt}
\end{equation}
Thus mean-field theory gives the correct scaling behaviour above the upper 
critical dimension $d_{\rm c}=2$. 

This picture is confirmed by exact results in one 
dimension (1d)~[4--6], rigourous 
bounds~\cite{bramson91}
and renormalization group results~\cite{peliti86,lee94} showing that the 
amplitude $C$ is universal. As usual, logarithmic corrections occur at $d_{\rm 
c}$ where the particle density decays 
as~\cite{lee94} 
\begin{equation}
\rho(t)=\frac{1}{8\pi}\,\frac{\ln t}{Dt}\,,\qquad d=d_{\rm c}=2\,,
\label{e1.ln}
\end{equation}
where $D$ is the diffusion constant.

We consider the influence of the decay exponent 
$\omega$, the amplitude $\lambda_0$ and the dimension $d$ of the system on 
the scaling behaviour of the particle density $\rho(t)$. In section~2, we 
give scaling arguments showing that the mean-field behaviour should be 
recovered when the decay exponent is greater than a critical value    
$\omega_{\rm c}(d)$ when $d\leq d_{\rm c}$ and calculate the time evolution of 
the particle density in mean-field theory. In section~3 these results are 
confronted with Monte Carlo simulations data in dimension $d=1$ to $3$. In 
section~4 we show how time-dependent reaction rates are effectively realized in 
the case of coupled reactions. Our results are 
summarized in section~5.

\section{Scaling considerations and mean-field theory}

In a continuum description in $d$ dimensions, the spacetime evolution of the 
particle density field, $\rho({\mathbf r},t)$, is governed by a Langevin 
equation containing diffusion, reaction and noise 
terms~[10--12]
\begin{equation}
\partial_t\rho({\mathbf r},t)=-\frac{\lambda_0}{t^\omega}\,\rho^2({\mathbf 
r},t)+D\,\nabla^2\rho({\mathbf r},t)+\zeta({\mathbf r},t)\,.
\label{e2.langevin}
\end{equation}
The noise term $\zeta({\mathbf r},t)$ accounts for the fluctuations of the 
particle density at position ${\mathbf r}$ at time $t$.

Let us consider the behaviour under rescaling of the reaction term
\begin{equation}
\partial_t\rho({\mathbf r},t)|_{\rm reaction}
=-\frac{\lambda_0}{t^\omega}\,\rho^2({\mathbf r},t)
\label{e2.rea1}
\end{equation}
at the stable fixed point of the system with a constant reaction rate, when 
the fluctuations are relevant, i.e., below the upper critical dimension 
$d_{\rm c}$.

Under a change of the length scale $L'=L/b$, the particle density field and 
its space average $\rho(t)$ scale with the same dimension $x_\rho$ and the 
scaling dimension of $t$ is the dynamical exponent $z=2$, so that
\begin{equation}
[\rho({\mathbf r},t)]'=b^{x_\rho}\rho({\mathbf r},t)\,,\quad 
t'=\frac{t}{b^z}\,.
\label{e2.scal1}
\end{equation}
For the average particle density, one obtains
\begin{equation}
[\rho(t)]'=\rho\left(\frac{t}{b^z}\right)=b^{x_\rho}\rho(t)
\label{e2.scal2}
\end{equation}
Taking $b=t^{1/z}$ leads to the power-law decay
\begin{equation}
\rho(t)=\rho(1)\,t^{-\alpha}\,,\qquad\alpha=\frac{x_\rho}{z}\,.
\label{e2.scal3}
\end{equation}

Using the transformations~(\ref{e2.scal1}) in~(\ref{e2.rea1}), we obtain
\begin{eqnarray}
\partial_{t'}[\rho({\mathbf r},t)]'|_{\rm reaction}
&=&b^{x_\rho+z}\partial_t\rho({\mathbf r},t)|_{\rm reaction}\nonumber\\
&=&-\frac{\lambda'_0}{{t'}^\omega}\,[{\rho}^2({\mathbf r},t)]'\nonumber\\
&=&-b^{z\omega+2x_\rho}\frac{\lambda'_0}{t^\omega}\,\rho^2({\mathbf r},t)
\label{e2.rea2}
\end{eqnarray}
so that the reaction-rate amplitude transforms as
\begin{equation}
\lambda'_0=b^{-z(\omega-1+x_\rho/z)}\lambda_0
=b^{-z(\omega-1+\alpha)}\lambda_0\,,
\label{e2.scal4}
\end{equation}
where the last expression follows from~(\ref{e2.scal3}).

When $\omega$ is smaller than the critical value $\omega_{\rm c}$, 
given by
\begin{equation}
\omega_{\rm c}=1-\alpha=1-\frac{d}{2}\,,\qquad d\leq d_{\rm c}\,,
\label{e2.omegac}
\end{equation}
according to~(\ref{e1.rhodt}), $\lambda_0$ increases under rescaling, i.e., 
the process is diffusion limited. The concentration fluctuations are relevant 
and the critical behaviour is governed by the same fixed point as for a 
constant reaction rate, for which $\alpha=d/2$.

When $\omega>\omega_{\rm c}$, the reaction-rate amplitude decreases and the 
process is reaction limited. The concentration fluctuations are suppressed by 
diffusion and the system should display a mean-field behaviour governed by a 
fixed line, parametrized by $\omega$. 

When $\omega=\omega_{\rm c}$, $\lambda_0$ is a marginal variable and 
logarithmic corrections to the mean-field behaviour are expected.

The behaviour of $\rho(t)$ follows from the 
mean-field rate equation
\begin{equation}
\dot{\rho}(t)
=-\lambda(t)\rho^2(t)
\label{e2.rea3}
\end{equation}
With the initial condition $\rho(t_0)=\rho_0$, the average density reads
\begin{eqnarray}
\frac{1}{\rho(t)}&=&\frac{1}{\rho_0}+\lambda_0\,
\frac{\left(t^{1-\omega}-t_0^{1-\omega}\right)}{1-\omega}\quad 
(\omega\neq1)\nonumber\\
\frac{1}{\rho(t)}&=&\frac{1}{\rho_0}
+\lambda_0\,\ln\left(\frac{t}{t_0}\right)\quad (\omega=1)
\label{e2.rho1}
\end{eqnarray}
In the asymptotic regime, $t\gg t_0$, one obtains
\begin{subeqnarray}
\rho(t)&\simeq&\frac{1-\omega}{\lambda_0}\,
t^{-(1-\omega)}\quad(\omega<1)\\
\rho(t)&\simeq&\frac{1}{\lambda_0\ln(t)}\quad(\omega=1)\\
\rho(t)&\simeq&\rho_\infty+\rho_\infty^2\,\lambda_0\,
\frac{t^{-(\omega-1)}}{\omega-1}\quad(\omega>1)
\label{e2.rho2}
\end{subeqnarray}
When $\omega>1$, the particle density decays algebraically to a non-vanishing 
asymptotic value, $\rho_\infty$, behaving as
\begin{eqnarray}
\frac{1}{\rho_\infty}&=&\frac{1}{\rho_0}+\lambda_0\,
\frac{t_0^{-(\omega-1)}}{\omega-1}\quad(\omega>1)\nonumber\\
\rho_\infty&\simeq&\frac{\omega-1}{\lambda_0}\qquad(\omega\to1+)
\label{e2.rho3}
\end{eqnarray}

When $d\geq d_{\rm c}$, $\omega_{\rm c}=0$ according to~(\ref{e2.omegac}). 
For positive values of $\omega$, the annihilation process is reaction limited 
and the mean-field behaviour in~(15a)--(15c) and~(\ref{e2.rho3}) applies.
For negative values of $\omega$ the reaction rate, which increases under 
rescaling, no longer controls the annihilation process. Thus one expect a 
scaling behaviour independent of $\omega$, the same as for $\omega=0$ with 
$\alpha=1$ and logarithmic corrections given by~(\ref{e1.ln}) at $d_{\rm c}$ 
when $\omega<0$.

\section{Monte Carlo simulations}

Numerical simulations of the diffusion-annihilation process have 
been performed in order to check the results of the last section for the 
asymptotic behaviour of the particle density $\rho(t)$ in dimensions $d=1$ to 
$3$. 

\subsection{Algorithm}
We work on hypercubic lattices with $L^d=10^6$ sites and periodic boundary 
conditions in all directions. In order to avoid finite-size effects the 
simulations are stopped when the diffusion length reaches some fraction 
($1/4$ to $1/10$) of the size $L$ of the system. The particle density 
$\rho(t)$ is averaged over 5 to 20 samples. 

At $t_0=1$ the $L^d$ sites are independently occupied by a particle with 
probability $\rho_0$. At time $t$, one of the $N(t)$ surviving  particles is 
randomly selected and a jump towards one of the 2d neighbouring sites is 
attempted, with the same probability for all the sites. When the target site 
is empty, the jump is accepted. When it is occupied, either the two particles 
annihilate with probability $\lambda(t)$ or they keep their original location 
with probability $1-\lambda(t)$, i.e., multiple occupancy of a site is not 
allowed. Finally, the time is incremented by $1/N(t)$ in all the cases and 
the process is repeated.

When $\omega<0$, the algorithm has to be modified since the reaction rate may 
increase beyond $1$. In this case, we let two particles annihilate with 
probability $\lambda(t)\tau(t)=1$. We take $\lambda_0=1$ such that 
$\lambda(t)\geq1$ and $\tau(t)\leq1$. The diffusion jumps are attempted with 
probability $\tau(t)\leq1$ and the time is incremented by $\tau(t)/N(t)$ at 
each Monte carlo step.

\subsection{Numerical results in 1d}
\begin{figure}[tbh]
\epsfxsize=7cm
\begin{center}
\vglue0.mm
\hspace*{-2.mm}\mbox{\epsfbox{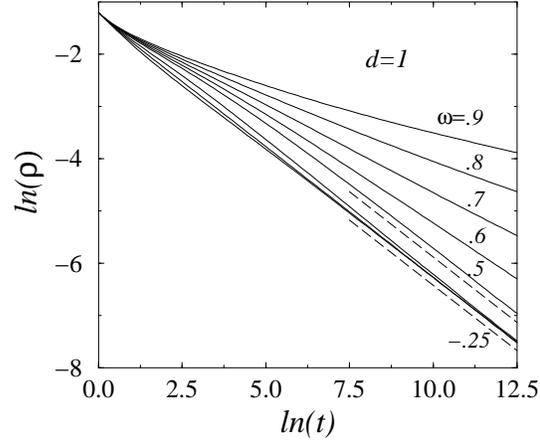}}
\end{center}
\vglue-5mm
\caption{Time dependence of the particle density in 1d for different values 
of the decay exponent $\omega$. The initial density is $\rho=0.3$ and the 
reaction-rate amplitude $\lambda_0=1$. The asymptotic slope, $-\alpha$, 
varies with $\omega$ in the reaction-limited mean-field regime, above the 
critical value $\omega_{\rm c}=1/2$. Below $\omega_{\rm c}$, in the 
diffusion-limited regime, it remains constant and equal to $-1/2$ (dashed 
lines). The amplitude is universal below $\omega_{\rm c}$, the data for 
$\omega=0.25$, $0$ and $-0.25$ collapsing asymptotically on a single line.}  
\label{fig1}  
\end{figure}
\begin{figure}[bht]
\epsfxsize=7cm
\begin{center}
\vglue0.mm
\hspace*{-2.mm}\mbox{\epsfbox{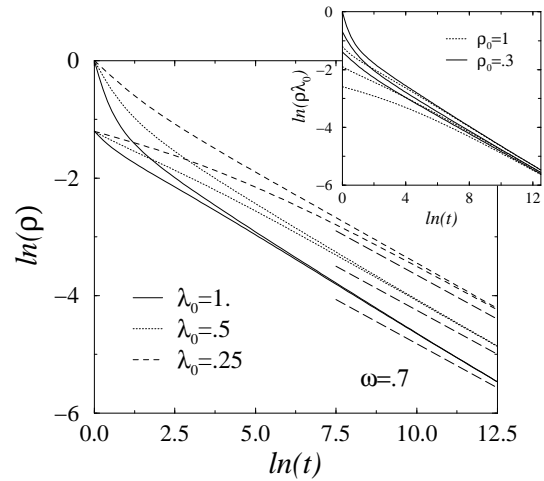}}
\end{center}
\vglue-5mm
\caption{Influence of $\rho_0$ and $\lambda_0$ on the amplitude of the 
particle density $\rho(t)$ in 1d. At $\omega=0.7$, i.e., in the 
reaction-limited regime, the amplitude does not depend on the initial density 
$\rho_0$ and it varies as $1/\lambda_0$ as shown in the inset where an 
asymptotical collapse is obtained for $\rho\lambda_0$.}  
\label{fig2}  
\end{figure}
\begin{figure}[tbh]
\epsfxsize=7cm
\begin{center}
\vglue0.mm
\hspace*{-2.mm}\mbox{\epsfbox{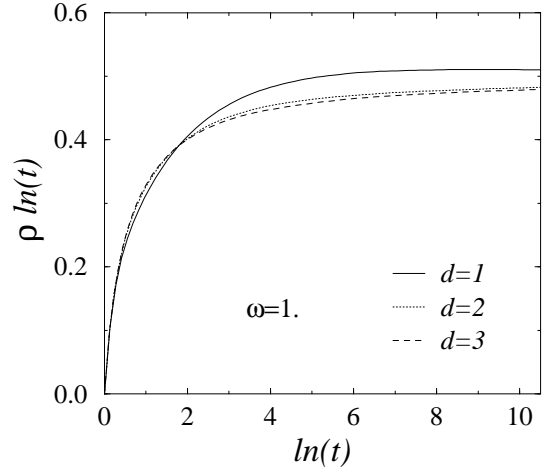}}
\end{center}
\vglue-5mm
\caption{Logarithmic scaling behaviour at $\omega=1$ for $d=1$ to $3$. In 
the mean-field reaction-limited regime the particle density decays 
asymtotically as $1/\ln(\rho)$ when $\omega=1$. The data were obtained with 
$\rho_0=1$ and $\lambda_0=1$}  
\label{fig3}  
\end{figure}
\begin{figure}[bht]
\epsfxsize=7cm
\begin{center}
\vglue0.mm
\hspace*{-2.mm}\mbox{\epsfbox{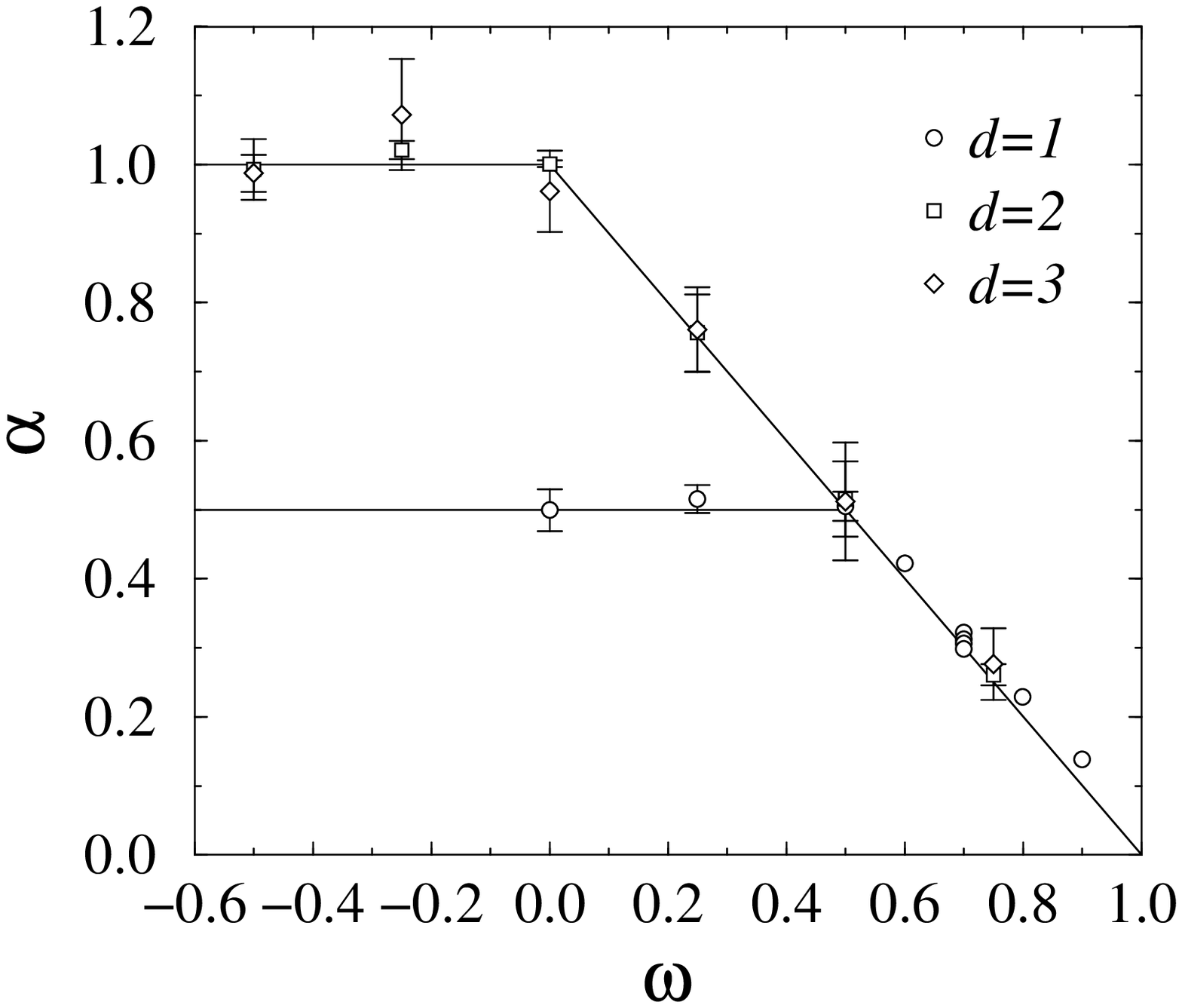}}
\end{center}
\vglue-5mm
\caption{Variation of the particle density exponent $\alpha$ with the 
reaction-rate exponent $\omega$ for $d=1,2$ and $3$. The data, obtained with 
$\rho=0.3$ and $\lambda_0=1$, are in overall agreement with the expected 
behaviour (solid lines). Below $\omega_{\rm c}$ ($1/2$ when $d=1$, $0$ when 
$d=2,3$) the process is diffusion limited and $\alpha$ is a constant. Above 
$\omega_{\rm c}$ the process is reaction limited and $\alpha$ decays as 
$1-\omega$ in agreement with mean-field theory.}  
\label{fig4}  
\end{figure}

The influence of $\omega$ on the scaling behaviour of $\rho(t)$ is shown in 
figure~\ref{fig1}. The asymtotic slope, $-\alpha$, is the same as for a 
constant reaction rate ($-1/2$, indicated by a dashed line) as long as 
$\omega\leq\omega_{\rm c}=1/2$. The amplitude of $\rho(t)$ is independent of 
$\omega$ below $\omega_{\rm c}$ and it is universal as for 
$\omega=0$~\cite{lee94}: it depends neither on the initial density $\rho_0$ 
nor on the reaction-rate amplitude $\lambda_0$, the process being 
diffusion limited. 

The form of the logarithmic correction to the algebraic decay, expected at 
$\omega_{\rm c}$ in relation with the marginality of $\lambda_0$, could not be 
extracted from our Monte Carlo data. 

Above $\omega_{\rm c}$, $\alpha$ decreases continuously to 0 when $\omega$ 
increases to 1. As shown in figure~\ref{fig2} the amplitude of $\rho(t)$ is 
no longer universal: it remains independent of $\rho_0$ but, the process 
being now reaction limited, it depends on $\lambda_0$. The inset shows that 
the amplitude actually varies as $1/\lambda_0$, in agreement with the 
mean-field result in~(\ref{e2.rho2}a).   

The scaling behaviour of $\rho(t)$ at $\omega=1$ is illustrated in 
figure~\ref{fig3}. The product $\rho\ln(t)$ tends to a constant value at long 
times, in agreement with mean-field theory in equation~(\ref{e2.rho2}b).

The variation of $\alpha$ with $\omega$ is illustrated in figure~\ref{fig4}. 
The exponents were deduced from the extrapolation of two-point approximants 
for the slope of $\ln(\rho)$ versus $\ln(t)$ using the Burlisch--Stoer (BS) 
algorithm~\cite{burlisch64,henkel88}. These data were obtained with 
$\rho_0=0.3$ in order to accelerate the approach of the asymptotic regime (see 
figure~\ref{fig2} for a comparison with $\rho_0=1$ when $\lambda_0=1$). The 
value $\lambda_0=1$ was selected to spare computer time although smaller 
values can lead to better estimates of the exponent as illustrated for 
$\omega=0.7$ where the circles correspond to values of $\lambda_0=1,0.4,0.2$ 
and $0.1$ from top to bottom. The results are in good agreement with the 
expected behaviour.

\subsection{Numerical results in 2d and 3d}

\begin{figure}[tbh]
\epsfxsize=7cm
\begin{center}
\vglue0.mm
\hspace*{-2.mm}\mbox{\epsfbox{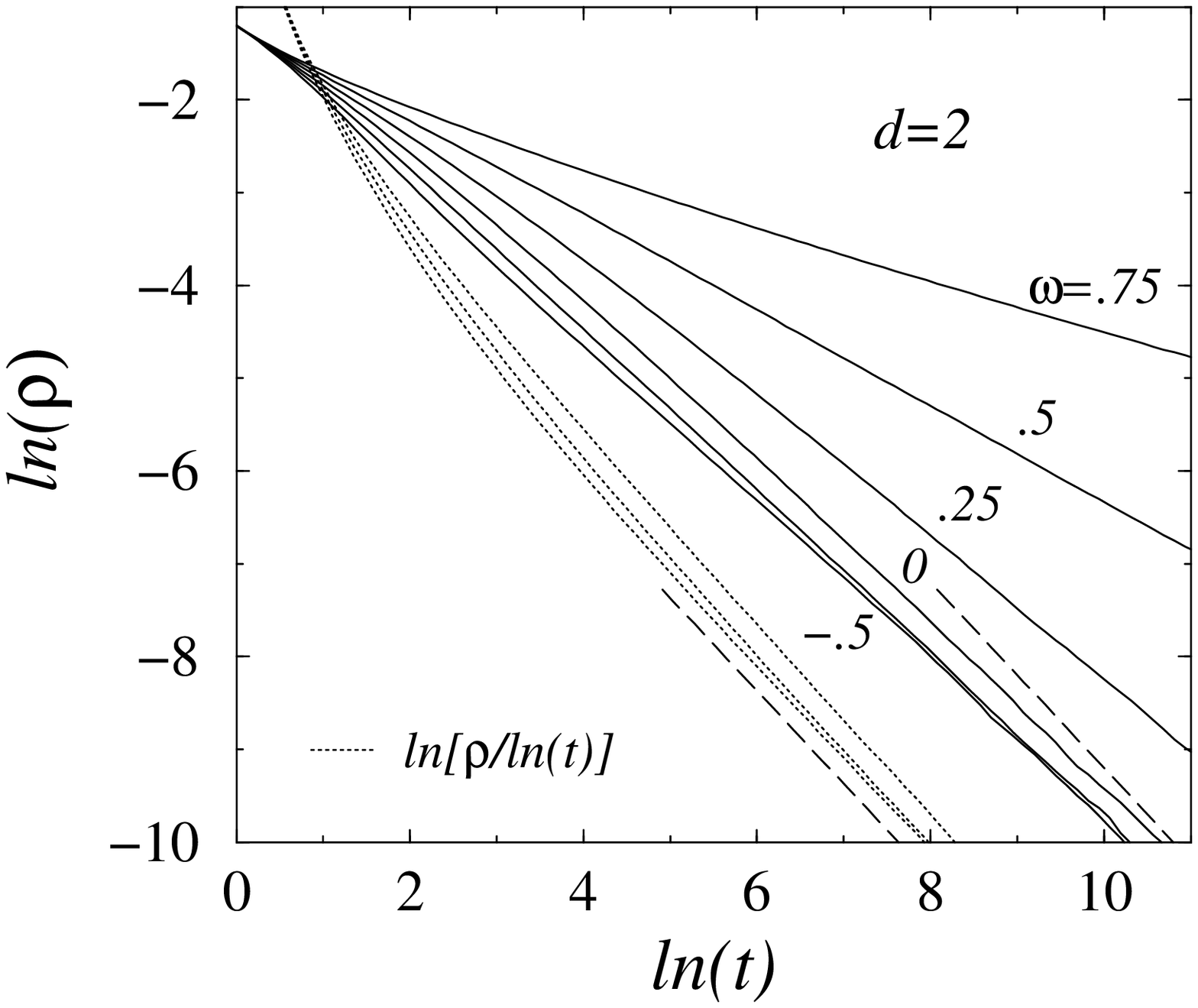}}
\end{center}
\vglue-5mm
\caption{Time dependence of the particle density in 2d with $\rho_0=0.3$ and 
$\lambda_0=1$. The asymptotic slope depends on $\omega$ in the reaction-limited 
regime $\omega>\omega_{\rm c}=0$ and remains constant in the diffusion-limited 
regime ($\omega=0,-0.25,-0.5$).  There is a deviation from the expected  
slope, $-1$ (dashed lines), due to the logarithmic correction occuring in the 
diffusion-limited regime at $d_{\rm c}=2$. When $\rho$ is divided by $\ln(t)$ 
(dotted lines) the correct slope is recovered.}  
\label{fig5}  
\end{figure}
\begin{figure}[hbt]
\epsfxsize=7cm
\begin{center}
\vglue0.mm
\hspace*{-2.mm}\mbox{\epsfbox{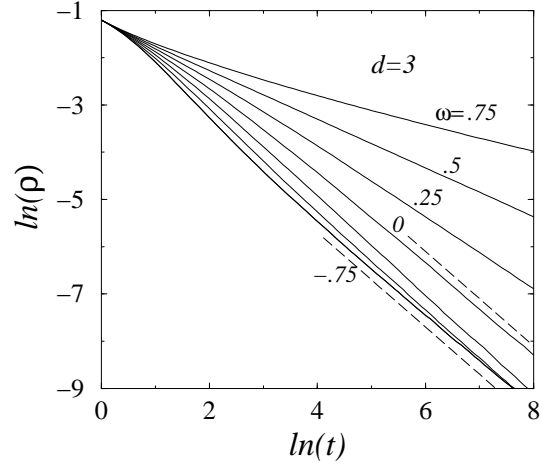}}
\end{center}
\vglue-5mm
\caption{Time dependence of the particle density in 3d with $\rho_0=0.3$ and 
$\lambda_0=1$. The critical value of the reaction-rate exponent is $\omega_{\rm 
c}=0$, the same as in 2d. Since $d>d_{\rm c}$, there are no logarithmic 
corrections and the asymptotic slopes are close to $-1$ (dashed lines) in the 
diffusion-limited regime ($\omega=0,-0.25,-0.5,-0.75$).}  
\label{fig6}  
\end{figure}

The log-log plots for the time evolution of the particle density for 
different values of $\omega$ are shown in figure~\ref{fig5} for $d=2$ and 
figure~\ref{fig6} for $d=3$. As above the data were obtained with an 
initial density $\rho_0=0.3$ and a rection-rate amplitude $\lambda_0=1$. In 
both cases the critical value of the reaction-rate exponent is 
$\omega_{\rm c}=0$. $\alpha$ varies with $\omega$ in the reaction-limited 
regime, $\omega>\omega_{\rm c}$, and remains constant in the diffusion-limited 
regime, $\omega\leq\omega_{\rm c}$. 

In 2d there is a systematic deviation from the slope $-1$ (indicated by a 
dashed line  in figure~\ref{fig5}) when $\omega\leq0$. This may be traced to 
the logarithmic correction in equation~(\ref{e1.ln}), occuring at the upper 
critical dimension $d_{\rm c}=2$ in the diffusion-limited regime. When $\rho$ 
is divided by $\ln(t)$ (dotted lines) the slopes are asymptotically close to 
the expected value $-1$ corresponding to the mean-field exponent $\alpha=1$. 
The amplitude is probably universal, the same for all values of 
$\omega\leq0$, although the evolution to the asymptotic behaviour becomes 
quite slow when approaching $\omega_{\rm c}=0$. 

In $d=3>d_{\rm c}$ the mean-field result of equation~(\ref{e2.rho2}a) with 
$\omega=0$, $\rho(t)\sim t^{-1}$, applies in the whole diffusion-limited 
regime $\omega\leq0$ as shown in figure~\ref{fig6} where the slopes are close 
to the expected value, indicated by the dashed lines, without any correction. 
Here too we expect the amplitude to be universal when $\omega\leq0$ but the 
evolution to the true asymptotic behaviour is even slower than in 2d when 
$\omega_{\rm c}$ is approached.

The asymptotic logarithmic decay of the particle density at $\omega\!=\!1$ 
(see equation~(\ref{e2.rho2}b)) is 
illustrated in figure~\ref{fig3}. The variation of $\alpha$ with $\omega$, 
shown in figure~\ref{fig4}, is still in agreement with the expected 
behaviour. The exponents were obtained through a non-linear least-square fit, 
taking into account an effective correction-to-scaling exponent, 
since the statistical fluctuations were too high to use the BS algorithm. 

\section{Time-dependent effective reaction rates}

\begin{figure}[tbh]
\epsfxsize=7cm
\begin{center}
\vglue0.mm
\hspace*{-2.mm}\mbox{\epsfbox{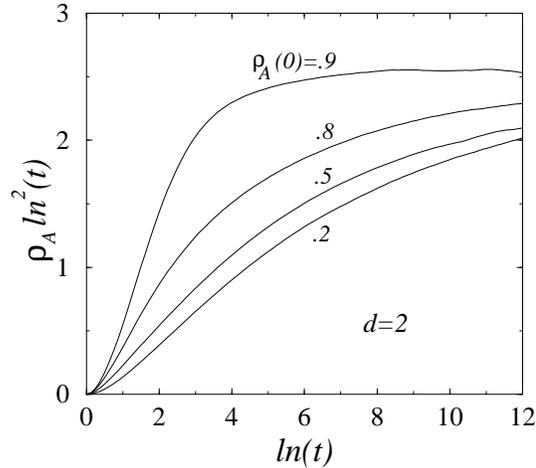}}
\end{center}
\vglue-5mm
\caption{Scaling behaviour of an asymmetric diffusion-annihilation process in 
2d where the annihilation of $A$ particles is catalysed by $B$ particles 
whereas the annihilation of $B$ particles is free. The product $\rho_A\ln^2(t)$ 
tends to a constant value at long time.}  
\label{fig7}  
\end{figure}
\begin{figure}[thb]
\epsfxsize=7cm
\begin{center}
\vglue0.mm
\hspace*{-2.mm}\mbox{\epsfbox{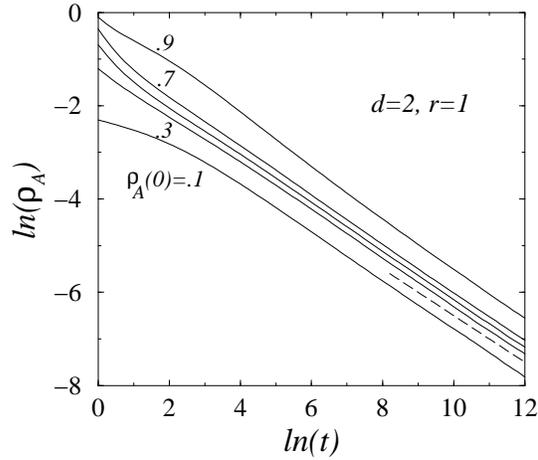}}
\end{center}
\vglue-5mm
\caption{Scaling behaviour of the particle density for a 2d symmetric 
diffusion-annihilation process where the annihilation of two particles 
of one species is catalysed by a particle of the other species. The reaction 
rates are equal, $r=\lambda_A/\lambda_B=1$, and both densities decay as 
$t^{-1/2}$. The density of $B$ particles with initial density $\rho_B(0)$ is 
equal to the density of $A$ particles with 
initial density $\rho_A(0)=1-\rho_B(0)$. The asymptotic slope, 
$-1/2$, is indicated by a dashed line.}  
\label{fig8}  
\end{figure}
\begin{figure}[hbt]
\epsfxsize=7cm
\begin{center}
\vglue0.mm
\hspace*{-2.mm}\mbox{\epsfbox{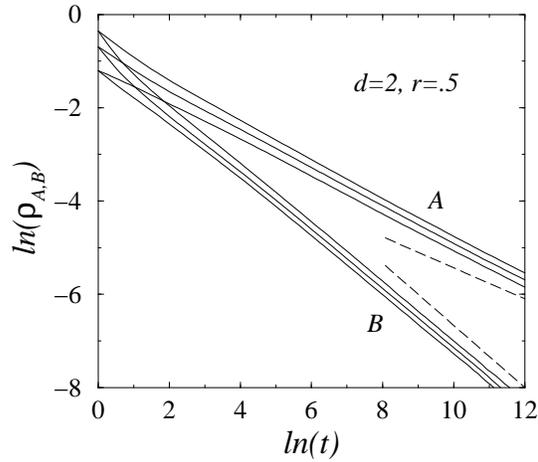}}
\end{center}
\vglue-5mm
\caption{Scaling behaviour of the particle density for a 2d symmetric 
diffusion-annihilation process where the annihilation of two particles 
of one species is catalysed by a particle of the other species. Here the 
reaction 
rates are different with $r=\lambda_A/\lambda_B=1/2$. The initial densities are 
such that $\rho_A(0)+\rho_B(0)=1$ and equal to $0.7$, $0.5$, $0.3$, from top to bottom 
for $A$ and $B$ particles. The expected asymptotic slopes, $-1/3$ for $A$ 
particles and $-2/3$ for $B$ particles, are indicated by dashed lines. The 
convergence is slow for the majority species.}  
\label{fig9}  
\end{figure}

Time-dependent effective reaction rates with a power-law decay can be realized 
in the case of coupled reactions. Let us first consider a  
diffusion-annihilation process in 2d where the annihilation of a pair of $A$ 
particles is catalysed by $B$ particles whereas $B$ particles annihilate 
without any further condition:  
\begin{eqnarray}
&&A\,B\,A\ \stackrel{\lambda_A}{\longrightarrow}\ \varnothing\,B\,\varnothing
\quad{\rm(catalysed\ annihilation)}\nonumber\\
&&B\,B\ \stackrel{\lambda_B}{\longrightarrow}\ \varnothing\,\varnothing
\quad{\rm(simple\ annihilation)}\,.
\label{e4.aba}
\end{eqnarray}

The effective reaction rate for $AA$ annihilation is governed by the mean 
density of $B$ particles, $\rho_B(t)$, if the concentration fluctuations are 
negligible, which is true in 2d. Thus $\omega_A=\alpha_B=1$ for the 
single-species annihilation in 2d. The $AA$ process being reaction limited 
with $\omega_A=1$, mean-field theory applies and a logarithmic decay is 
expected. Actually, one has to take into account the logarithmic correction in 
equation~(\ref{e1.ln}) so that the effective reaction rate is 
$\lambda_A(t)=\lambda_{A0}\ln(t)/t$. The 
mean-field rate equation~(\ref{e2.rea3}) leads to the asymptotic behaviour
\begin{equation}
\rho_A(t)\simeq\frac{2}{\lambda_{A0}}\frac{1}{\ln^2(t)}\,.
\label{e4.ln}
\end{equation}

This process has been simulated on the square lattice with 
$\lambda_A=\lambda_B=1$ and $\rho_A(0)+\rho_B(0)=1$. The size of the system, 
the 
boundary conditions and the number of samples are the same as before. When a 
particle jump is attempted towards an occupied site, the two particles 
annihilate in the case of a $B$ pair. For a pair of $A$ particles, the 
annihilation occurs only when a $B$ particle is first neighbour of one of the 
$A$ particles and second neighbour of the second. When the two particles are 
different they just keep their positions.

The Monte Carlo results, shown in figure~7, confirm this scaling behaviour 
although the convergence is quite slow for small or intermediate initial 
densities of $A$ particles. 
The second example concerns a system of two symmetrically coupled 
single-species diffusion-annihilation processes in 2d with
\begin{eqnarray}
&&A\,B\,A\ \stackrel{\lambda_A}{\longrightarrow}\ \varnothing\,B\,\varnothing
\quad{\rm(catalysed\ annihilation)}\nonumber\\
&&B\,A\,B\ \stackrel{\lambda_B}{\longrightarrow}\ \varnothing\,A\,\varnothing
\quad{\rm(catalysed\ annihilation)}\,.
\label{e4.abab}
\end{eqnarray}
The density of one species controls the reaction rate of the other so that 
$\omega_A=\alpha_B$. Now, assuming that the system 
is in the reaction-limited regime, equation~(\ref{e2.rho2}a) leads to 
$\alpha_A=1-\omega_A=1-\alpha_B$ and
\begin{equation}
\alpha_A+\alpha_B=1\,.
\label{e5.abab}
\end{equation}
When the process is fully symmetric, i.e., when $\lambda_A=\lambda_B$, both 
species decay with the same exponent $\alpha_A=\alpha_B=1/2$, a 
value consistent with the assumed reaction-limited mean-field behaviour 
since $\omega_{\rm c}=0$ in 2d. Actually, the fully symmetric 
process reduces to the  $3A\ {\rightarrow}\ A$ process which is known to decay 
as $t^{-1/2}$ above its upper critical dimension $d_{\rm 
c}=1$~\cite{benavraham93}.

The Monte Carlo simulations on the square lattice with 
$r=\lambda_A/\lambda_B=1$, $\rho_A(0)+\rho_B(0)=1$ and the same rules as above 
for the catalysed annihilation of the two species are shown in figure~8. The 
scaling behaviour is in complete agreement with a decay exponent equal to 
$1/2$. 

A deviation of the reaction-rate ratio $r$ from 1 is sufficient to break the 
symmetry between the two species, leading to different decay exponents. This 
follows from the solution of the system of coupled mean-field rate equations
\begin{equation}
\dot{\rho}_A(t)
=-\lambda_A\rho_A^2(t)\rho_B(t)\,,\qquad
\dot{\rho}_B(t)
=-\lambda_B\rho_B^2(t)\rho_A(t)\,.
\label{e6.abab}
\end{equation}
Multiplying the first equation by $\rho_B$, the second by $\rho_A$ and adding, 
one obtains a differential equation for $\rho_A\rho_B$ leading to the 
asymptotic behaviour 
\begin{equation}
\rho_A(t)\rho_B(t)\sim t^{-1}
\label{e7.abab}
\end{equation}
in agreement with~(\ref{e5.abab}). 
Dividing the first equation by $\rho_A$ and using~(\ref{e7.abab}), one finally 
obtains
\begin{equation}
\rho_A(t)\sim t^{-r/(1+r)}\,,\qquad\rho_B(t)\sim t^{-1/(1+r)}\,,\qquad 
r=\lambda_A/\lambda_B\,.
\label{e8.abab}
\end{equation}
The decay exponents continuously vary with the ratio $r$ of the reaction 
rates but their sum remains equal to 1 as expected from the effective 
reaction-rate argument leading to~(\ref{e5.abab}). Actually, it is easy 
to verify that $\lambda_A$ and $\lambda_B$ are marginal variables when 
$\alpha_A+\alpha_B=1$. 

The Monte Carlo results are shown in figure~9 for $d=2$ and $r=1/2$. They 
confirm this behaviour, although the convergence to the asymptotic slope is 
rather slow for the majority species. 

\section{Conclusion}
We have shown, through scaling considerations and Monte Carlo simulations, that 
when the reaction rate of the single-species diffusion-annihilation process 
decays as $t^{-\omega}$, there is a critical value 
$\omega_{\rm c}=1-d/2$ for $d\leq d_{\rm c}$ separating reaction-limited 
behaviour for $\omega>\omega_{\rm c}$ from diffusion-limited behaviour for  
$\omega<\omega_{\rm c}$. In the reaction-limited regime, mean-field theory is 
always valid and leads to an $\omega$-dependent decay of the particle density 
whereas in the diffusion-limited regime, due to concentration fluctuations, the 
particle density decays as $t^{-d/2}$ when $d\leq d_{\rm c}=2$ as for a 
constant reaction rate. 

In the case of coupled reactions where the annihilation of one type of particles 
is catalysed by the other, one obtains a time-dependent effective reaction rate 
which is easy to identify when the process is reaction limited, i.e., when 
mean-field theory applies. 

The main effect of decreasing time-dependent reaction rates is to extend 
the validity of mean-field theory below the usual upper critical dimension and 
to lead to unusual mean-field behaviour.

\ack

The Laboratoire de Physique des Mat\'eriaux is Unit\'e Mixte de 
Recherche CNRS no~7556.


\Bibliography{99}

\bibitem{kang84a} Kang K and Redner S 1984 \PR A {\bf 30} 2833

\bibitem{toussaint83} Toussaint D and Wilczek F 1983 \JCP {\bf 78} 2642

\bibitem{kang85} Kang K and Redner S 1985 \PR A {\bf 32} 435

\bibitem{lushnikov86} Lushnikov A A 1986 {\it Sov. Phys.--JETP} {\bf 64} 811

\bibitem{spouge88} Spouge J L 1988 \PRL {\bf 60} 871

\bibitem{benavraham90} ben-Avraham D, Burschka M A and Doering C R 1990 {\it J. 
Stat. Phys.} {\bf 60} 695

\bibitem{bramson91} Bramson M and Lebowitz J L 1991 {\it J. Stat. Phys.} {\bf 
62} 297

\bibitem{peliti86} Peliti L 1986 \JPA {\bf 19} L365

\bibitem{lee94} Lee B P 1994 \JPA {\bf 27} 2633

\bibitem{cardy98} Cardy J 1998 Field theory and non\-equilibrium 
sta\-tis\-tical me\-chanics (Troisi\`eme cy\-cle de la phys\-ique en Su\-isse 
Ro\-mande) {\it Preprint}
http:\-//www\--thphys.\-physics\-.ox.ac.uk\-/users\-/JohnCardy\-/notes.ps

\bibitem{hinrichsen00} Hinrichsen H 2000 {\it Adv. Phys.} {\bf 49} 815

\bibitem{odor02} \'Odor G 2004 \RMP {\bf 76} 663

\bibitem{burlisch64} Burlisch R and Stoer J 1964 {\it Numer. Math.} {\bf 6} 413

\bibitem{henkel88} Henkel M and Sch\"utz G 1988 \JPA {\bf 217} 2617

\bibitem{benavraham93} ben-Avraham D 1993 \PRL {\bf 71} 3733

\endbib
\end{document}